\documentclass[epj]{svjour}

\usepackage{latexsym}
\usepackage{graphicx}
\usepackage{fancyhdr}
\usepackage{cite,amsmath,amssymb,bbm}

\def\makeheadbox{{
 }}
\def\mytitle{My title} 
\def\myauthors{My name}  
\def\mytype{My type of session}
\def\mysession{My session}

\def\mytitle{Moduli stabilization in (string) model building:
gauge fluxes and loops}
\def\myauthors{Michele Trapletti}    
\def\mytype{Contributed Talk}    
\def\mysession{Theoretical Models}
\newcommand{\be}{\begin{equation}}  
\newcommand{\ee}{\end{equation}}  
\newcommand{\bea}{\begin{eqnarray}}  
\newcommand{\eea}{\end{eqnarray}}

\newcommand{\ol}[1]{\overline{#1}}
\newcommand{\ind}{\operatorname{ind}}

\begin{document}

\title{Moduli stabilization in (string) model building:\\
gauge fluxes and loops}
\author{
A.~P.~Braun\inst{1}
\and A.~Hebecker\inst{1}
\and M.~Trapletti\inst{2}$^,$\inst{3}
} 

\institute{
Institut f\"ur Theoretische Physik, Universit\"at Heidelberg,
Philosophenweg 16 und 19, D-69120 Heidelberg, Germany.
\and 
Laboratoire de Physique
Theorique, Bat. 210, Universit\'e de Paris-Sud,
F-91405 Orsay, France. 
\and
Centre de Physique Th\'eorique, \'Ecole Polytechnique,
F-91128 Palaiseau, France.
}

\date{}

\abstract{
\noindent
We discuss the moduli stabilization arising in the presence of gauge fluxes,
R--symmetry twists and non--perturbative effects in the context of 6-dimensional 
supergravity models.
We show how the presence of $D$-terms, due to the gauge fluxes, is compatible
with gaugino condensation, and that the two effects, combined with the R--symmetry
twist, do stabilize all the K\"ahler moduli present in the model, in the spirit of KKLT.
We also calculate the flux-induced one-loop correction to the scalar potential coming
from charged hypermultiplets, and find that it does not destabilize the minimum.
} 

\maketitle

\section{Introduction}
One of the perceived problems of the KKLT
construction~\cite{Kachru:2003aw}, is the presence of $\ol{\rm D3}$-branes
(`anti-D3-branes'), which break SUSY explicitly and do not have a
supergravity description.
\footnote{
It has, however, been argued that a phenomenologically motivated description in 
terms of non-linearly realized supersymmetry is sufficient for most practical 
purposes~\cite{Choi:2005ge}. Indeed,
when modelling the $\ol{\rm D3}$ brane effect by $F$-term breaking, the 
phenomenology turns out to be independent of the detailed dynamics of this 
SUSY breaking sector (unless extra fields violate the underlying sequestering
assumption)~\cite{Brummer:2006dg}.}

Following \cite{Burgess:2003ic}, one could avoid such a problem by trading
the $\ol{\rm D3}$-branes  for the introduction of two-form flux on the worldvolume
of D7-branes, that has a supergravity description in terms of a SUSY-breaking
$D$-term potential  (see e.g.~\cite{Kachru:1999vj,Jockers:2005zy,Villadoro:2006ia,
Haack:2006cy}).
Unfortunately, there are two fundamental problems with this proposal:
one related to the intimate connection between $F$- and  $D$-terms,
the other to the gauge invariance of the superpotential~\cite{
Dudas:2005vv,Choi:2005ge,Villadoro:2005yq,deAlwis:2006sz}.
Namely, the $D$-terms originate from the gauging of an isometry of the scalar
manifold of the supergravity model. In KKLT, such an isometry should act on the single
K\"ahler modulus  $T$ by shifting its imaginary part. This clashes with the
fact that the superpotential  $W=W_0+A e^T$ is not invariant under such a shift.
This clash can obviously be avoided if light fields other than $T$ are 
present \cite{Dudas:2005vv,Kashani-Poor:2005si,Achucarro:2006zf,Haack:2006cy},
but in this case a reanalysis of the whole stabilization/uplifting proposal is needed.

In our following investigation we approach the problem from a
6d supergravity perspective\cite{Salam:1984cj,Aghababaie:2002be,gibbons},
in the presence of 2-form-flux. 

In Sect.~2 we introduce a $T^2/Z_{2}$ model
in which two moduli superfields $S$ and $T$ encode the dilaton and the
compactification volume. We calculate the scalar potential arising in the presence of 
2-form-flux in two ways: by integrating the $F_{56}^2$ term over the compact 
space and by finding the $D$-term that arises from the gauge transformation of 
$T$. Since the superfield $S$, which governs all gauge-kinetic 
functions, does not transform, no gauge invariance problem arises in the 
presence of gaugino condensation (see also \cite{Haack:2006cy}).

In Sect.~3 we introduce Scherk-Schwarz SUSY breaking as a source for a 
constant superpotential $W_0$. We study the compatibility of such an option with
a $T^2/Z_n$ orbifold compactification, finding that
only the $n=2$ case is actually viable. 

In Sect.~4 we calculate the one-loop correction to the scalar
potential. This is done by the explicit computation of the correction
that arises if hypermultiplets charged under the fluxed U(1) are present.
Since the constituents of  the charged hypermultiplet feel the flux directly,
we expect this to be the dominant contribution to the corrections.

In Sect.~5 we discuss options for moduli stabilization using the various 
ingredients analysed above. Working on a $T^2/Z_2$ orbifold and ignoring, 
for simplicity, the shape modulus of the torus, one still has to deal with 
the stabilization of the superfields $S$ and $T$ simultaneously. At fixed $T$, 
the modulus $S$ is stabilized \`a la KKLT by the interplay of $W_0$ and 
gaugino condensate. The depth of the resulting SUSY AdS vacuum depends on 
$T$, driving Re$T$ to small values. This is balanced by the $T$
dependence of the flux-induced $D$-term, leading to a stable non-SUSY AdS 
vacuum. Thus, while the 2-form flux does not provide the desired uplift,
it plays an essential role in the simultaneous stabilization of two moduli. 
Finally, we show that the loop-corrections do not spoil the stabilization.
\vspace{-12pt}
\paragraph{Acknowledgements:} The work of MT is supported by the European
Community through the contract 041273.

\section{A six-dimensional model: gauge fluxes as a source for $D$-term potential}
We start from the following bosonic action for supergravity coupled to gauge
theory in six dimensions~\cite{Nishino:1986dc,Salam:1984cj}
(for details see \cite{Braun:2006se} and references therein)
\begin{equation}
\mathcal{L}=
-\frac{\sqrt{-g_6}}{2}\left[\mathcal{R}_{6}+\left(\partial
\phi\right)^2+\frac{e^{2\phi}}{12}H^2
+\frac{e^{\phi}}{2}F^2\right], \label{L}
\end{equation}
where
$H\equiv dB+F\wedge A$.
This action is invariant under the gauge transformations
\be
\delta A=d\Lambda, \,,\,\,\,\,\,\,
\delta B=-\Lambda F+dC\,.\label{Btrafo}
\ee
We consider a compactification on $\mathbb{M}^4\times T^2$, with
\begin{equation}
\label{metric}
{g_{6}}_{MN}=
  \left(\begin{array}{cc}
  r^{\mbox{}\hspace{-3pt}-2}{g_{4}}_{\mu\nu } &
  \mbox{}\hspace{-12pt}0 \\
  0 & \mbox{}\hspace{-12pt} r^{2}{g_{2}}_{mn}
  \end{array}\right)\hspace{-1pt},\,\,
{g_{2}}_{mn}\hspace{-4pt}=
\displaystyle\frac{1}{\tau_{2}}
        \left(\begin{array}{cc} 
	1 & \tau_{1} \\
	\tau_{1} & |\tilde\tau|^2 
	\end{array}\right)\hspace{-1pt},
\end{equation}
where $\mu,\nu = 0 .. 3$, $m,n= 5 .. 6$, and
$\tilde\tau\equiv\tau_{2}+i\tau_{1}$.
The domain of $x_{5}$ and $x_{6}$ is taken to be a square of unit 
length, so that $\int\sqrt{g_{2}}d\hspace{-.2ex}x^{5}d\hspace{-.2ex}x^{6}=1$. 

We introduce a constant background for the field strength $\langle F_{mn} 
\rangle=f\epsilon_{mn}$, $f$ being a quantized number.
We split the gauge potential $A$ into a fluctuation term 
$\mathcal A$ and a background term $\langle A\rangle$, such that $\langle 
F\rangle=d\langle A\rangle$. The background $\langle A\rangle$ cannot be 
globally defined in the internal space, thus, also $B$ is not globally defined.
Rather, a new field $\mathcal B=B-\langle A\rangle \wedge \mathcal A$ is
globally defined~\cite{Kaloper:1999yr} and has a standard Kaluza-Klein
expansion.

The next step is to pass to the 4d theory arising from the compactification on a
supersymmetric $T^2/Z_2$ orbifold. 
We achieve this by disregarding all 4d vector multiplets arising from 6d gravity,
as well as the Wilson line degrees of freedom associated with the 6d gauge
theory (we neglect the possibility of localized matter). 
What remains is 4d supergravity, the $A_\mu$ vector multiplet,
and three chiral multiplets. The latter contain the degrees of freedom 
$r$, $\phi$, $\tilde\tau$ and two scalars related to the 
2-form $\mathcal B$. The lowest components of the three modulus superfields 
are~\cite{Aghababaie:2002be,Falkowski:2005zv}
\begin{equation}
S\equiv\tfrac{1}{2}(s+ic),\quad T\equiv\tfrac{1}{2}(t+ib),
\quad \tau\equiv\tfrac{1}{2}(\tau_{2}+i\tau_{1})\label{defTStau}.
\end{equation}
where we have used the definitions
$ t\equiv e^{-\phi}r^{2}$, $s\equiv e^{\phi}r^{2}$
and $b\epsilon_{mn}\equiv \mathcal B_{mn}$,
$\epsilon_{\mu\nu\rho\sigma} \partial^{\sigma}c\equiv 
r^{4} e^{2\phi}(d\mathcal B)_{\mu\nu\rho}$.
The K\"ahler potential, which can be inferred from the kinetic terms for the 
scalars, is
\begin{equation} 
K=-\log(T+\overline{T})-\log(S+\overline{S})-\log(\tau+\overline{\tau})\,.
\end{equation} 
Similarly, the gauge-kinetic function is found to be $h(S)=2S$.

The 4d model is invariant under the gauge transformations inherited from
those described in Eq.~(\ref{Btrafo}). In particular, the gauge transformations
of ${\cal B}$ follow from its definition together with Eq.~(\ref{Btrafo}).
Considering 4d gauge transformations $\Lambda=\Lambda(x^\mu)$ 
and restricting to the zero-mode level only, we find
$\delta  \mathcal B_{56}=-2 \Lambda \langle F_{56}\rangle$, i.e.
$\delta b=-2f\Lambda$.
This implies that the only nonvanishing component of the Killing vector
is $X^{T}=-if$. Thus, the resulting $D$-term, $D=iK_TX^T$, leads to the
$D$-term potential
\be
V_{D}=f^{2}/2st^{2}.
\ee
The same potential also follows directly from the 6d gauge-kinetic
term, evaluated in the flux background.
This represents a nontrivial check of the fact that the flux is described by 
the gauging of an isometry from the 4d perspective. (See~\cite{Villa} 
for a similar computation in heterotic string theory.)
Note in particular that the gauge transformation acts only on $T$, while the
gauge kinetic function depends only on $S$.
Hence, no clash between gaugino condensate and $D$-term
potential arises. A related situation occurring in the presence of both flux
and gaugino  condensation on the {\it same} D7-brane-stack has been
discussed  in~\cite{Haack:2006cy}.

\section{Scherk-Schwarz twists as a source for $W_0$}
The 6d supergravity theory studied in Sect.~2 possesses an SU(2)$_{\rm R}$
R-symmetry, thus we can compactify it on $T^2$ imposing non-trivial
field-identifications. Given a generic SU(2)$_{\rm R}$ doublet 
$\Phi(x^\mu,x^5,x^6)$ (e.g. the gaugino) we require
\bea
\Phi(x^\mu,x^5,x^6)=T_5 \Phi(x^\mu,x^5+1,x^6),\\
\Phi(x^\mu,x^5,x^6)=T_6 \Phi(x^\mu,x^5,x^6+1),
\eea
where the matrices $T_i$ embed the translations $t_i$ along the torus
coordinate $x^i$ in the R-symmetry group. Since $t_5 t_6=t_6 t_5$, we also 
require $T_5 T_6=T_6 T_5$. In case one (or both) of the matrices are 
non-trivial, we obtain a Scherk-Schwarz (SS)  dimensional reduction \cite{SS}.

For an orbifold compactification of the 6d theory, the rotation
operator $r\!\!\in\,$SO(2) is also embedded in the R-symmetry group via a 
matrix $R$. A non-trivial embedding is crucial to avoid a hard SUSY breaking,
indeed, in case $R={\mathbbm 1}$, the net action of the orbifold on any 4d
spinor would result in a non-trivial phase, projecting it out of the spectrum. 
Having such a non-trivial embedding, extra consistency conditions must be 
fulfilled.

In the case of a $Z_2$ orbifold, $r^2=1$, $r t_i=t_i^{-1} r$, and we have
to impose these conditions also on the corresponding transformations of the 
spinors. Non-trivial solutions to these conditions exist~\cite{Lee:2005tk}, 
as can be easily demonstrated explicitly: The transformation associated with 
$r$ is $\tilde{R}=S(r)R$, where $S(r)$ is the phase rotation of the two 4d 
Weyl spinors coming from a {\bf 4} of SO(1,5). In the $Z_2$ case, we have 
$S(r)=i{\mathbbm 1}$. Choosing $R=\mbox{diag}(-i,i)$, we find $\tilde{R}= 
\mbox{diag}(1,-1)$.
This matrix satisfies the required consistency relations 
with $T_i=\exp(i\alpha_i\sigma_2)$. In case only one of the $T_i$'s is non 
trivial, e.g. $\alpha_6=0$ and $\alpha_5=\alpha$, we can shrink the $x^6$ 
direction, obtaining a 5d effective field theory compactified on $S^1/Z_2$. 
In this case it is well known that the continuous SS parameter $\alpha$
can be described by a tunable constant superpotential $W_0\sim\alpha$~\cite{
Dudas:1997jn}. In the rest of the paper, we mainly consider such a $T^2/Z_2$ 
compactification, the 4d field content of which was already anticipated in 
Sect.~2. Notice that with such a field content a constant $W_0$ leads, in 
absence of any other effects, to SUSY breaking with zero tree-level 
potential, as expected in a SS reduction.

In case of a $Z_3$, $Z_4$ or $Z_6$ reduction, the field content would be even 
more appealing, since the $\tau$ multiplet would be projected away. However,
the  consistency conditions for a SS reduction are now more stringent and
cannot  be satisfied.

\section{Loop corrections in the 6d model}
In order to estimate the loop corrections in the presence of flux, we consider the
one-loop Casimir energy of a charged 6d hypermultiplet. We expect this to be the
dominant contribution since the constituents of the charged hypermultiplet feel the
flux directly. Moreover,
since flux quantization implies quantized coefficient for this loop correction,
we also expect that the latter will be more important than the Casimir
energy induced by all the other (weak) SUSY breaking effects.
In this sense, we expect the computation worked out here to
provide a good estimate of the total loop corrections to our model.

We first derive the mass 
spectra of the charged 6d scalars and Weyl fermions (see \cite{Bachas:1995ik}). 
A 6d hypermultiplet  consists of two complex scalars and one 6d Weyl fermion
which enter the action in a quite complicated way~\cite{Nishino:1986dc}.
We will linearize the  $\sigma$-model and work with canonical kinetic terms,
neglecting the  self-interactions of the scalars. This is expected to be a good 
approximation as long as the mass scale of gauge interactions in 6d is much
lower than the 6d~Plank~scale, $1/g_{\rm YM,6}\ll M_{\rm Pl,6}$.

The masses of the scalars are given by the  eigenvalues of the Laplacian
on the compact space. For one minimally coupled complex scalar field with
covariant derivative $\mathcal{D}$, the Laplacian reads
$r^{-4}\left(\mathcal{D}_{5}^{2}+\mathcal{D}_{6}^{2}\right)$,
where we have used the decomposition of Eq.~(\ref{metric}), assuming
$\tau_1=0$ and $\tau_2=1$. 
In the case of a nonzero constant flux the covariant derivatives no longer
commute:
$\left[\mathcal{D}_{5},\mathcal{D}_{6}\right]=iF_{56}=if$.
Algebraically, this is equivalent to a one-dimensional harmonic oscillator
with unit mass and unit frequency. 
For positive $f$ we can identify the ``position'' operator with $\mathcal{D}_{5}$
and the ``momentum'' operator with $\mathcal{D}_{6}$, for negative $f$, the
position and momentum operators have to be interchanged
but the mass spectrum is not affected. It reads
\be
m_{n}^{2}=\frac{2\vert f\vert}{r^{4}}\left(n+\tfrac{1}{2}\right),\,\,\,n>0.
\label{mbos}
\ee

The masses of the fermions are instead given by
\begin{equation}
m_{n}^{2}r^{4}\psi_{n}=
\left(\Gamma^{5}\mathcal{D}_{5}+
\Gamma^{6}\mathcal{D}_{6}\right)^{2}\psi_{n},
\end{equation}
where the $\psi_{n}$ are 6d spinors. 
Since
\begin{equation}
\left(\Gamma^{5}\mathcal{D}_{5}+\Gamma^{6}\mathcal{D}_{6}\right)^{2}=
\mathcal{D}_{5}^{2}
+\mathcal{D}_{6}^{2}+i\Gamma^{5}\Gamma^{6}f,
\end{equation}
the problem differs from the bosonic case only by a shift,
if the spinors are eigenvectors of $\Gamma^{5}\Gamma^{6}$.
Since $\Gamma^{7}=i\gamma^{5}\Gamma^{5}\Gamma^{6}$,
and the 6d chirality is fixed, such a shift is given by 
the chirality of each 4d spinor obtained by decomposing the 6d spinor.
The mass spectrum of 4d 
Weyl fermions reads 
\begin{equation}
(m_{n}^{2})_{\pm}=\frac{2\vert f\vert}{r^{4}}
\left(n+\tfrac{1}{2}\pm\tfrac{1}{2}\right).\label{mferm}
\end{equation} 

Given the spectra, we need to find their degeneracy.
By using the index theorem we find that
\begin{equation}
\ind(\Gamma^{5}\mathcal{D}_{5}+\Gamma^{6}\mathcal{D}_{6})=\frac{1}{2\pi}
\int_{T^{2}}F=\frac{f}{2\pi}=N\label{index}.
\end{equation}
Thus the monopole number equals the degeneracy of the state with vanishing 
mass. It is clear that the ground state of the fermions of opposite chirality 
has the same degeneracy, because we are considering the same Laplace
operator to which merely a constant is added, and thus we find precisely
the same eigenfunctions.
By the same argument we conclude that the bosonic ground state is $N$-fold
degenerate.

With this particle spectrum we directly compute the one-loop effective
potential from a four-dimensional perspective (see \cite{Braun:2006se} for
details), finding
\be
V_{C}=
\frac{7}{4}\frac{\vert N\vert^{3}}{(st)^{2}}\zeta_{R}'(-2)
\cong-\frac{0.053}{(2\pi)^{3}}\frac{\vert f\vert^{3}}{(st)^{2}}\label{VC-no-z2}   
\ee
Here we have used the quantization condition for the flux, Eq.~(\ref{index}).
\par

The computation is analogous, albeit technically more involved, 
in the $T^{2}/Z_{2}$ case. The result is: 
\bea
V_{C}^{\pm}=7\frac{N^2}{(st)^2} \zeta'(-2)J_{N}^{\pm}
\cong-\frac{0.053}{(2\pi)^{2}}\frac{f^2}{(st)^2}J_{N}^{\pm},
\eea
where we have defined
\be
J_{N}^{\pm}\equiv \vert N\vert\pm 4.
\ee
The two signs in $V^{\pm}$ stem from the different
internal parity that may be assigned to the fermions on the
massless level. 

This correction can be understood as a correction to the 
K\"ahler potential.
We found a non-zero Casimir energy because SUSY is broken, which in
turn is a result of the flux. The flux was shown to generate a $D$-term
potential in  Sect.~2. We can trace the correction to the $D$-term potential
back to a correction to the K\"ahler potential if we assume that the gauge 
symmetries of our model remain unchanged. Neglecting higher orders in
$1/r$ we find 
\be
\frac{f^{2}}{st}(\Delta K)_{T}
=-\frac{1}{(2\pi)^{2}}\frac{7}{4}\zeta'(-2)
\frac{f^2}{(st)^2}J_{N}^{\pm},
\ee
so that we can conclude
\be
\Delta K=-\frac{1}{(2\pi)^{2}}\frac{7}{4}
\frac{\zeta'(-2)}{S+\ol{S}}\log(T+\ol{T})J_{N}^{\pm}.
\ee

\section{Moduli stabilization}
In this section we study the stabilization of our model. 
We start by considering the effect of a gaugino condensate, a constant
superpotential term and the $D$-term due to the gauge flux.
We neglect for a moment the contribution due to the loop corrections.
We have
\begin{eqnarray}
K &\hspace{-3pt}=&\hspace{-3pt} -\log(T+\overline{T})-\log(S+\overline{S})-\log(\tau+\overline{\tau}),\\
W &\hspace{-3pt}=&\hspace{-3pt} \mu^{3}\exp(-aS)+W_{0},
\end{eqnarray} 
where we assume for simplicity that $a$ and $\mu$ are real and positive,
and $W_{0}$ is real and negative.
The complete scalar potential, including the $D$-term is then
\be
V=
\frac{\tilde{V}(s)}{t(\tau+\overline{\tau})}+\frac{f^{2}}{2st^{2}},
\ee
with
\be \label{VV}
\tilde{V}(s)=
a\mu^{6}(as+2)e^{-as}\hspace{-3pt}+
2W_{0}\mu^{3}a\cos\left(\frac{ac}{2}\right)e^{-\frac{as}{2}}
\ee
This potential stabilizes both $s$ and $t$ at a negative value of $V$, as is
shown in the following.

Consider first the `axionic' partner of $s$, denoted by $c$. As $W_{0}$ is 
taken to be negative, while $a$ and $\mu^{3}$ are positive, $c$ is always 
stabilized at a value where the cosine is unity. Thus we assume $c=0$ in the 
following. Since the shift symmetry acting on the modulus $b$  (the `axionic'
partner of $t$) is gauged, $b$ is absorbed in the massive vector boson. 
Further effects have to be taken into account to stabilize the complex 
structure modulus $\tau$, for which we assume $2\tau=1$ from now on.

To get some intuition for the stabilization of $s$ and $t$, it is 
advantageous to first set $f=0$ and $t=1$. Then the remaining modulus 
$s$ enters the potential in exactly the same fashion as in the KKLT model. 
At the minimum of the potential, $s$ has to solve $D_{S}W=0$, so that we find
$W_{0}+\mu^{3}e^{-\frac{as}{2}}(1+as)=0$.
This is equivalent to minimizing $\tilde{V}(s)$. For small $W_{0}$ we find the
approximate solution
\begin{equation}
as_{0}\sim 2\ln(-\mu^{3}/W_{0}).
\end{equation} 
This equation shows that $as_{0}$ can be made parametrically large by tuning 
$W_{0}$ to have small negative values.

The approximate value at which $t$ is stabilized can be found by setting 
$s=s_{0}$. This is reasonable as the extra 
$1/s$ contribution coming from the $D$-term potential will not alter the value
of $s$ at the minimum significantly. The resulting potential for $t$ is then 
\begin{equation}
V(t)=\frac{f^{2}}{2s_{0}t^{2}}+\frac{\tilde{V}(s_{0})}{t},\label{Vt}
\end{equation}
which is minimized by $t_{0}=-f^{2}/s_{0}\tilde{V}(s_{0})$.

The perturbative corrections of Sect.~3 do not alter the stabilization
qualitatively.   As a contribution to the effective action, they can simply be added
to the scalar potential, and slightly drive the minimum to larger values of $s$ and $t$.

\end{document}